\documentclass[prd,aps,groupedaddress,showpacs]{revtex4}
\usepackage{epsf,epsfig,graphicx}

\newcounter{muni}

\begin{document}
\hbadness=10000 \pagenumbering{arabic}

\title{Puzzles in $B$ physics}

\author{Hsiang-nan Li}
\email{hnli@phys.sinica.edu.tw}

\affiliation{Institute of Physics, Academia Sinica, Taipei, Taiwan
115, Republic of China,} \affiliation{Department of Physics,
National Cheng-Kung University, Tainan, Taiwan 701, Republic of
China}

\begin{abstract}

I discuss some puzzles observed in exclusive $B$ meson decays,
concentrating on the large difference between the direct CP
asymmetries in the $B^0\to \pi^\mp K^\pm$ and $B^\pm\to \pi^0
K^\pm$ modes, the large $B^0\to\pi^0\pi^0$ branching ratio, and
the large deviation of the mixing-induced CP asymmetries in the
$b\to sq\bar q$ penguins from those in the $b\to c\bar c s$ trees.

\end{abstract}


\maketitle

\section{Introduction}

The $B$ factories have accumulated enough events, which allow
precision measurements of exclusive $B$ meson decays. These
measurements sharpen the discrepancies between experimental data
and theoretical predictions within the standard model, such that
some puzzles have appeared. The recently observed direct CP
asymmetries and branching ratios of the $B\to \pi K$, $\pi\pi$
decays \cite{HFAG},
\begin{eqnarray}
A_{CP}(B^0\to \pi^\mp K^\pm)&=&(-10.8\pm 1.7)\%\;,\nonumber\\
A_{CP}(B^\pm\to \pi^0 K^\pm)&=&(4\pm 4)\%\;,\nonumber\\
B(B^0\to\pi^\mp\pi^\pm)&=&(4.9\pm 0.4)\times 10^{-6}\;,\nonumber\\
B(B^0\to\pi^0\pi^0)&=&(1.45\pm 0.29)\times 10^{-6}\;,\label{data}
\end{eqnarray}
are prominent examples. The expected relations $A_{CP}(B^0\to
\pi^\mp K^\pm)\approx A_{CP}(B^\pm\to \pi^0 K^\pm)$ and
$B(B^0\to\pi^\mp\pi^\pm)\gg B(B^0\to\pi^0\pi^0)$ obviously
contradict to the above data. The weak phase $\phi_1$, defined via
the Cabibbo-Kobayashi-Maskawa (CKM) matrix element
$V_{td}=|V_{td}|\exp(-i\phi_1)$ \cite{KM}, can be extracted either
from the tree-dominated or penguin-dominated modes. It has been
estimated that the penguin pollution in the $b\to c\bar cs$ trees
and the tree pollution in the $b\to sq\bar q$ penguins are about
5\%. Therefore, it is expected that the measured mixing-induced CP
asymmetries $S_{sq\bar q}$ are close to $S_{c\bar
cs}=\sin(2\phi_1)\approx 0.685$ \cite{HFAG}. However, a large
deviation $\Delta S\equiv S_{sq\bar q}-S_{c\bar cs}$ has been
measured.

In this talk I will review the recent studies of these subjects,
concluding that the $B\to\pi K$ puzzle could be attributed to QCD
uncertainty, the $B\to\pi\pi$ puzzle can not be resolved within
the current theoretical development, and the $\Delta S$ puzzle
might be a promising signal of new physics, if the data persist. I
will not discuss another puzzle from the small longitudinal
polarization fractions observed in the penguin-dominated $B\to VV$
decays, such as $B\to\phi K^*$ and $B\to\rho K^*$, since they
involve different dynamics. A recent summary on this topic is
referred to \cite{M06}.

\section{The $B\to\pi K$ puzzle}

To explain the $B\to\pi K$ puzzle, it is useful to adopt the
topological-amplitude parametrization for two-body nonleptonic $B$
meson decays \cite{CC}. The $B\to \pi K$ amplitudes are written,
up to $O(\lambda^2)$, $\lambda\approx 0.22$ being the Wolfenstein
parameter, as
\begin{eqnarray}
A(B^+\to \pi^+ K^0)&=&P'\;,\nonumber\\
\sqrt{2}A(B^+\to \pi^0 K^+)&=&-P'\left[1+\frac{P'_{ew}}{P'}
+\left(\frac{T'}{P'}+\frac{C'}{P'} \right)e^{i\phi_3}\right]\;,
\nonumber\\
A(B^0\to \pi^-
K^+)&=&-P'\left(1+\frac{T'}{P'}e^{i\phi_3}\right)\;,
\nonumber\\
\sqrt{2}A(B^0\to \pi^0K^0)&=&P'\left(1 -\frac{P'_{ew}}{P'}
-\frac{C'}{P'}e^{i\phi_3}\right)\;. \label{Mbpp1}
\end{eqnarray}
The notations $T'$, $C'$, $P'$, and $P'_{ew}$ stand for the
color-allowed tree, color-suppressed tree, penguin, and
electroweak penguin amplitudes, respectively, which obey the
counting rules \cite{GHL,Charng},
\begin{eqnarray}
\frac{T'}{P'}\sim  \lambda\;,\;\;\;\;
\frac{P'_{ew}}{P'}\sim\lambda\;,\;\;\;\;
\frac{C'}{P'}\sim\lambda^2\;. \label{pow}
\end{eqnarray}
The weak phase $\phi_3$ is defined via the CKM matrix element
$V_{ub}=|V_{ub}|\exp(-i\phi_3)$ \cite{KM}. The data $A_{CP}(B^0\to
\pi^\mp K^\pm)\approx -11\%$ indicate a sizable relative strong
phase between $T'$ and $P'$, which verifies our prediction made
years ago using the perturbative QCD (PQCD) approach \cite{KLS}.
Since both $P_{ew}'$ and $C'$ are subdominant, the approximate
equality for the direct CP asymmetries $A_{CP}(B^\pm\to
\pi^0K^\pm)\approx A_{CP}(B^0\to \pi^\mp K^\pm)$ is expected,
which is, however, in conflict with the data in Eq.~(\ref{data})
dramatically.

It is then natural to conjecture a large $P_{ew}'$
\cite{Y03,BFRPR,NK04,GR03,CFMM}, which signals a new physics
effect, a large $C'$ \cite{Charng2,HM04,CGRS,Ligeti04}, which
implies a missing mechanism in the standard model, or both
\cite{WZ,BHLD}. The large $C'$ proposal seems to be favored by a
recent analysis of the $B\to \pi K$, $\pi\pi$ data based on the
amplitude parametrization \cite{Charng2}. The PQCD predictions for
the $B\to \pi K$, $\pi\pi$ decays in \cite{KLS,LUY} were derived
from the leading-order (LO) and leading-power formalism. While LO
PQCD gives a negligible $C'$, it is possible that this supposedly
tiny amplitude receives a significant subleading correction.
Hence, before claiming a new physics signal, one should at least
examine whether the next-to-leading-order (NLO) effects could
enhance $C'$ significantly.

\begin{figure}[htbp]
\epsfxsize=7cm \centerline{\epsfbox{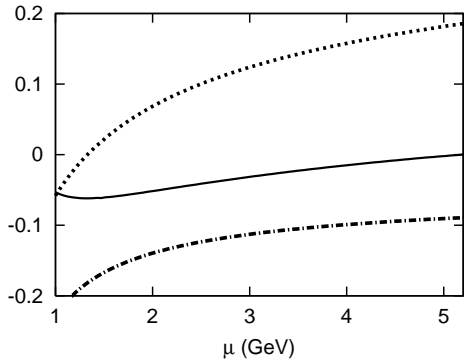}} \caption{Real part of
$a_2$ for the $B\to \pi K$ decays without the vertex corrections
(dotted lines) and with the vertex corrections (solid lines), and
imaginary part with the vertex corrections (dot-dashed lines) in
the NDR scheme.} \label{fig:vc}
\end{figure}

In \cite{LMS} we calculated the important NLO contributions to the
$B\to \pi K$, $\pi\pi$ decays from the vertex corrections, the
quark loops, and the magnetic penguins. Those NLO corrections to
the $B$ meson transition form factors, being overall quantities,
are irrelevant. The higher-power corrections, having not yet been
under good control, were not considered. We found that the
corrections from the quark loops and from the magnetic penguins,
being about 10\% of the LO penguin amplitude, decrease only the
$B\to\pi K$ branching ratios as shown in Table~\ref{br1}. The
vertex corrections increase $C'$ by a factor of 3, and induce a
large phase relative to $T'$. This result can be understood from
the value of the Wilson coefficient $a_2(\mu)$ in
Fig.~\ref{fig:vc}, to which $C'$ is proportional, at the
characteristic scale $\mu\approx \sqrt{m_b\Lambda}\approx 1.7$
GeV, $m_b$ being the $B$ quark mass and $\Lambda$ a hadronic
scale. The larger $C'$ renders the total tree amplitude $T'+C'$
more or less parallel to the total penguin amplitude $P'+P'_{ew}$
in the $B^\pm\to \pi^0K^\pm$ modes. Hence, it leads to nearly
vanishing $A_{CP}(B^\pm\to \pi^0K^\pm)$ as shown in
Table~\ref{cp1}, and the $B\to \pi K$ puzzle is resolved at the
$1\sigma$ level. Our analysis also confirmed that the NLO
corrections are under control in PQCD.

\begin{table}[ht]
\begin{tabular}{cccccccc}
\hline Mode & Data \cite{HFAG}& LO & LO$_{\rm NLOWC}$ & +VC & +QL
& +MP & +NLO
\\
\hline $B^\pm \to \pi^\pm K^0$ & $ 24.1 \pm 1.3 $ &
 $17.0$&$32.3$&$30.1$&$34.2$&$24.1$&
 $23.6^{+14.5\,(+13.8)}_{-\ 8.4\,(-\ 8.2)}$
\\
$B^\pm \to \pi^0 K^\pm$ & $ 12.1 \pm 0.8 $ &
 $10.2$&$18.4$&$17.1$&$19.4$&$14.0$&
 $13.6^{+10.3\,(+\ 7.3)}_{-\ 5.7\,(-\ 4.3)}$
\\
$B^0 \to \pi^\mp K^\pm$ & $ 18.9 \pm 0.7 $ &
 $14.2$&$27.7$&$26.1$&$29.4$&$20.5$&
 $20.4^{+16.1\,(+11.5)}_{-\ 8.4\,(-\ 6.7)}$
\\
$B^0 \to \pi^0 K^0 $ & $ 11.5 \pm 1.0 $ &
 $\phantom{0}5.7$&$12.1$&$11.4$&$12.8$&$\phantom{0}8.7$&
 $\phantom{0}8.7^{+\ 6.0\,(+\ 5.5)}_{-\ 3.4\,(-\ 3.1)}$
\\
\hline $B^0 \to \pi^\mp \pi^\pm$ & $ \phantom{0}4.9 \pm 0.4 $ &
 $\phantom{0}7.0$&$\phantom{0}6.8$&$\phantom{0}6.6$&
 $\phantom{0}6.9$&$\phantom{0}6.7$&
 $\phantom{0}6.5^{+\ 6.7\,(+\ 2.7)}_{-\ 3.8\,(-\ 1.8)}$
\\
$B^\pm \to \pi^\pm \pi^0$ & $ \phantom{0}5.5 \pm 0.6 $ &
 $\phantom{0}3.5$&$\phantom{0}4.1$&$\phantom{0}4.0$&
 $\phantom{0}4.1$&$\phantom{0}4.1$&
 $\phantom{0}4.0^{+\ 3.4\,(+\ 1.7)}_{-\ 1.9\,(-\ 1.2)}$
\\
$B^0 \to \pi^0 \pi^0$ & $ \phantom{0}1.45 \pm 0.29 $ &
 $\phantom{0}0.12$&$\phantom{0}0.27$&$\phantom{0}0.37$&
 $\phantom{0}0.29$&$\phantom{0}0.21$&
 $\phantom{0}0.29^{+0.50\,(+0.13)}_{-0.20\,(-0.08)}$
\\\hline
\end{tabular}
\caption{Branching ratios from PQCD in the NDR scheme in units of
$\times 10^{-6}$. The label LO$_{\rm NLOWC}$ means the LO results
with the NLO Wilson coefficients, and +VC, +QL, +MP, and +NLO mean
the inclusions of the vertex corrections, of the quark loops, of
the magnetic penguin, and of all the above NLO corrections,
respectively. The theoretical uncertainties in the parentheses
represent those only from the variation of hadronic
parameters.}\label{br1}
\end{table}

\begin{table}[ht]
\begin{tabular}{cccccccc}
\hline Mode & Data \cite{HFAG}& LO & LO$_{\rm NLOWC}$& +VC & +QL &
+MP & +NLO
\\
\hline $B^\pm \to \pi^\pm K^0$ & $ -2 \pm  4$ &
 $\ \, -1$&$-1$&$-1$& $\phantom{-}0$&$\ \, -1$&
 $\ \ \ \ \, \phantom{-}0\pm 0\,(\pm 0)$
\\
$B^\pm \to \pi^0 K^\pm$ & $ \phantom{-}4 \pm 4 $ &
 $\ \, -8$&$-6$&$-2$&$-5$&$\ \, -8$&
 $\ \, -1^{+3\,(+3)}_{-6\,(-5)}$
\\
$B^0 \to \pi^\mp K^\pm$ & $ -10.8 \pm 1.7 $ &
 $-12$&$-8$&$-9$&$-6$&$-10$&
 $-10^{+7\,(+5)}_{-8\,(-6)}$
\\
$B^0 \to \pi^0 K^0 $ & $ \phantom{-}2 \pm 13 $ &
 $\ \, -2$& $\phantom{-}0$&$-7$& $\phantom{-}0$&
 $\ \, \phantom{-}0$&
 $\ \, -7^{+3\,(+1)}_{-4\,(-2)}$
\\
\hline $B^0 \to \pi^\mp \pi^\pm$ & $ \phantom{-}37 \pm 10$ &
 $\phantom{-}14$& $\phantom{-}19$& $\phantom{-}21$&
 $\phantom{-}16$& $\phantom{-}20$&
 $\phantom{-}18^{+20\,(+\ 7)}_{-12\,(-\ 6)}$
\\
$B^\pm \to \pi^\pm \pi^0$ & $ \phantom{-}1 \pm 6 $&
 $\ \, \phantom{-}0$& $\ \, \phantom{-}0$& $\ \, \phantom{-}0$&
 $\ \, \phantom{-}0$& $\ \, \phantom{-}0$&
 $\ \ \ \ 0\pm 0\,(\pm 0)$
\\
$B^0 \to \pi^0 \pi^0$ & $ \phantom{-}28^{+40}_{-39}$ &
 $\ \, -4$&$-34$& $\phantom{-}65$&$-41$&$-43$&
 $\phantom{-}63^{+35\,(+\ 9)}_{-34\,(-15)}$
\\\hline
\end{tabular}
\caption{Direct CP asymmetries from PQCD in the NDR scheme in
percentage.}\label{cp1}
\end{table}

At last, we emphasize that the NLO PQCD predictions for the
$B^0\to\pi^0 K^0$ still fall short a bit compared to the data. It
implies a new-physics phase associated with the electroweak
penguin amplitude $P'_{ew}$ \cite{BFRPR,HNS,BCLL,LY05}, such that
it becomes orthogonal to the penguin amplitude $P'$, and enhances
the $B^0\to\pi^0 K^0$ branching ratio. That is, we can not exclude
the possibility of new physics effects in the $B\to\pi K$ decays.

\section{The $B\to\pi\pi$ puzzle}

Similarly, the $B\to\pi\pi$ decay amplitudes are parameterized as
\begin{eqnarray}
\sqrt{2}A(B^+\to \pi^+\pi^0)&=&-T\left[1+\frac{C}{T}
+\frac{P_{ew}}{T}e^{i\phi_2}\right]\;,
\label{a1}\\
A(B_d^0\to \pi^+\pi^-)&=&-T\left(1
+\frac{P}{T}e^{i\phi_2}\right)\;,
\label{b1}\\
\sqrt{2}A(B_d^0\to \pi^0\pi^0)&=&T\left[\left(
\frac{P}{T}-\frac{P_{ew}}{T}\right) e^{i\phi_2}-\frac{C}{T}
\right]\;, \label{bpi1}
\end{eqnarray}
with the power counting rules,
\begin{eqnarray}
\frac{P}{T}\sim\lambda\;,\;\;\;\; \frac{C}{T}\sim
\lambda\;,\;\;\;\;\frac{P_{ew}}{T}\sim \lambda^2\;. \label{po}
\end{eqnarray}
The hierarchy of the branching ratios $B(B^0\to\pi^0\pi^0)\sim
O(\lambda^2)B(B^0\to\pi^\mp\pi^\pm)$ is then expected. However,
the data in Eq.~(\ref{data}) show $B(B^0\to\pi^0\pi^0)\sim
O(\lambda)B(B^0\to\pi^\mp\pi^\pm)$, giving rise to the
$B\to\pi\pi$ puzzle.

As indicated in Table~\ref{br1}, the NLO corrections, despite of
increasing the color-suppressed tree amplitudes significantly, are
not enough to enhance the $B^0\to\pi^0\pi^0$ branching ratio to
the measured value. A much larger amplitude ratio $|C/T|\sim 0.8$
must be obtained in order to resolve the puzzle \cite{Charng2}.
Nevertheless, the NLO corrections do improve the consistency of
our predictions with the data: the predicted
$B^0\to\pi^\pm\pi^\mp$ ($B^0\to\pi^0\pi^0$) branching ratio
decreases (increases). To make sure the NLO effects observed in
Sec.~2 are reasonable, we have applied the same PQCD formalism to
the $B\to\rho\rho$ branching ratios \cite{LM06}, which are also
sensitive to the color-suppressed tree contribution. It was found
that the NLO PQCD predictions are in agreement with the data of
the $B^0\to\rho^\mp\rho^\pm$ and $B^\pm\to\rho^\pm\rho^0$
branching ratios, and saturate the experimental upper bound of the
$B^0\to\rho^0\rho^0$ branching ratio as shown in Table~\ref{br2}.
We conclude that it is unlikely to accommodate the measured
$B^0\to\pi^0\pi^0$, $\rho^0\rho^0$ branching ratios simultaneously
in PQCD. Therefore, our resolution to the $B\to\pi K$ puzzle makes
sense, and the $B\to\pi\pi$ puzzle is confirmed.

\begin{table}[ht]
\begin{tabular}{ccccccccc}
\hline Mode & Data \cite{HFAG}  & LO & LO$_{\rm NLOWC}$ & +VC &
+QL & +MP & +NLO
\\
\hline $B^0 \to \rho^\mp \rho^\pm$ & $ 25.2^{+3.6}_{-3.7} $ &
$27.8$& $26.1$&$25.2$&$26.6$&$25.9$&
$25.3^{+25.3\,(+12.1)}_{-13.8\,(-\ 7.9)}$
\\
$B^\pm \to \rho^\pm \rho^0$ & $19.1\pm 3.5$ & $13.7$&
$16.2$&$16.0$&$16.2$&$16.2$& $16.0^{+15.0\,(+\ 7.8)}_{-\ 8.1\,(-\
5.3)}$
\\
$B^0 \to \rho^0 \rho^0$ & $ <1.1 $ &
$0.33$&$0.56$&$1.02$&$0.62$&$0.45$&
$0.92^{+1.10\,(+0.64)}_{-0.56\,(-0.40)}$
\\\hline
\end{tabular}
\caption{$B\to\rho\rho$ branching ratios from PQCD in the NDR
scheme in units of $10^{-6}$.}\label{br2}
\end{table}

It has been claimed that the $B\to\pi\pi$ puzzle is resolved in
the QCD-improved factorization (QCDF) approach \cite{BBNS} with an
input from soft-collinear effective theory (SCET) \cite{BY05}: the
inclusion of the NLO jet function, the hard coefficient of
SCET$_{\rm II}$, into the QCDF formula for the color-suppressed
tree amplitude gives sufficient enhancement of the
$B^0\to\pi^0\pi^0$ branching ratio. It is certainly necessary to
investigate whether the new mechanism proposed above deteriorates
the consistency of theoretical results with other data. Therefore,
we have extended the formalism in \cite{BY05} to the
$B\to\rho\rho$ decays as a check \cite{LM06}. Because of the
end-point singularities present in twist-3 spectator amplitudes
and in annihilation amplitudes, these contributions have to be
parameterized in QCDF \cite{BBNS}. Different scenarios for
choosing the free parameters, labelled by ``default", ``S1",
``S2", $\cdots$, and ``S4", have been proposed in \cite{BN}. As
shown in Table~\ref{tab1}, the large measured $B^0\to\pi^0\pi^0$
branching ratio can be accommodated by including the NLO jet
function, when the parameter scenario S4 is adopted. However, this
effect overshoots the upper bound of the $B^0\to\rho^0\rho^0$
branching ratio very much. We have surveyed the other scenarios,
and found the results from S1 and S3 (S2) similar to those from
the default (S4). That is, it is also unlikely to accommodate the
$B\to\pi\pi$, $\rho\rho$ data simultaneously in QCDF.

\begin{table}[ht]
\begin{tabular}{cccccc}
\hline Mode&Data \cite{HFAG} & default, LO jet  & default, NLO jet
& S4, LO jet & S4, NLO jet
\\\hline
$B^\pm\to\pi^\pm\pi^0$& $ \phantom{0}5.5 \pm 0.6 $ &
\phantom{0}6.02 & \phantom{0}6.24  & \phantom{0}5.07  &
\phantom{0}5.77
\\
$B^0\to\pi^\mp\pi^\pm$& $ \phantom{0}4.9 \pm 0.4 $ &
\phantom{0}8.90  & \phantom{0}8.69  & \phantom{0}5.22 &
\phantom{0}4.68
\\
$B^0\to\pi^0\pi^0$& $ \phantom{0}1.45 \pm 0.29 $ & \phantom{0}0.36
 & \phantom{0}0.40 & \phantom{0}0.72   &
\phantom{0}1.07
\\
$B^\pm\to\rho^\pm_L\rho^0_L$&$19.1\pm 3.5$ & 18.51 &19.48 &16.61 &
18.64
\\
$B^0\to\rho^\mp_L\rho^\pm_L$&$25.2^{+3.6}_{-3.7}$ & 25.36 & 24.42
& 18.48 & 16.76
\\
$B^0\to\rho^0_L\rho^0_L$&$<1.1$ &
\phantom{0}0.43&\phantom{0}0.66&\phantom{0}0.92&\phantom{0}1.73
\\\hline
\end{tabular}
\caption{Branching ratios from QCDF with the input of the SCET jet
function in units of $10^{-6}$. The data for the $B\to\rho\rho$
decays include all polarizations.}\label{tab1}
\end{table}

There exists an alternative phenomenological application of SCET
\cite{BPRS,BPS05}, where the jet function, characterized by the
scale of $O(\sqrt{m_b\Lambda})$, is regarded as being
incalculable. Its contribution, together with other
nonperturbative parameters, such as the charming penguin, were
then determined by the $B\to\pi\pi$ data. That is, the
color-suppressed tree amplitude can not be explained, but the data
are used to fit for the phenomenological parameters in the theory.
Predictions for the $B\to\pi K$, $KK$ decays were then made based
on the obtained parameters and partial SU(3) flavor symmetry
\cite{BPS05}. Final-state interaction (FSI) is certainly a
plausible resolution to the $B\to\pi\pi$ puzzle, but the estimate
of its effect is quite model-dependent. Even opposite conclusions
were drawn sometimes. When including FSI either into naive
factorization \cite{CHY} or into QCDF \cite{CCS}, the
$B^0\to\pi^0\pi^0$ branching ratio was treated as an input in
order to fix the involved free parameters. Hence, no resolution
was really proposed. It has been found that FSI, evaluated in the
Regge model, is insufficient to account for the observed
$B^0\to\pi^0\pi^0$ branching ratio \cite{DLLN}.

\section{The $\Delta S$ puzzle}

The time-dependent CP asymmetry of the $B^0\to \pi^0K_S$ mode is
defined as
\begin{eqnarray}
A_{CP}(B^0(t)\to \pi^0K_S) &\equiv& \frac{B({\bar B}^0(t)\to
\pi^0K_S)- B(B^0(t)\to \pi^0K_S)} {B({\bar B}^0(t)\to
\pi^0K_S)+B(B^0(t)\to \pi^0K_S)}
\nonumber\\
&=& A_{\pi^0K_S}\cos(\Delta M_d\, t)+S_{\pi^0K_S}\sin(\Delta M_d\,
t)\;, \label{CPk}
\end{eqnarray}
with the mass difference $\Delta M_d$ of the two $B$-meson mass
eigenstates, and the direct asymmetry and the mixing-induced
asymmetry,
\begin{eqnarray}
A_{\pi^0K_S}\, =\, {|\lambda_{\pi^0K_S}|^2-1 \over
1+|\lambda_{\pi^0K_S}|^2}\;, \hspace{20mm} S_{\pi^0K_S}\, =\,
{2\,\rm{Im}(\lambda_{\pi^0K_S}) \over
1+|\lambda_{\pi^0K_S}|^2}\;,\label{spk}
\end{eqnarray}
respectively. The $B^0\to\pi^0 K_S$ decay has a CP-odd final
state, and the corresponding factor,
\begin{eqnarray}
\lambda_{\pi^0 K_S}\, =\, -e^{-2i\phi_1} {P' -P'_{ew}
-C'e^{-i\phi_3} \over P' -P'_{ew} -C'e^{i\phi_3}} \;.\label{mix}
\end{eqnarray}

After obtaining the values of the various topological amplitudes,
we computed the mixing-induced CP asymmetries through
Eq.~(\ref{mix}) \cite{LMS}. Since $C'$ is of $O(\lambda^2)$
compared to $P'$, it is expected that the LO PQCD result of
$S_{\pi^0K_S}\approx 0.70$ is close to $S_{c\bar cs}\approx 0.685$
as shown in Table~\ref{mixcp}. It is known that the leading
deviation of $\Delta S_{\pi^0 K_S}\equiv S_{\pi^0K_S}-S_{c\bar
cs}$ caused by $C'$ is proportional to
$\cos(\delta_{C'}-\delta_{P'})$, if neglecting $P'_{ew}$, where
$\delta_{C'}$ ($\delta_{P'}$) is the strong phase of $C'$ ($P'$).
Because the vertex corrections induce a large $\delta_{C'}$, $C'$
becomes more orthogonal to $P'$, and $\Delta S_{\pi^0 K_S}$ does
not increase much in NLO PQCD. This tendency persists in other
$b\to sq\bar q$ penguin decays. The mixing-induced CP asymmetry in
the $B^0\to\pi^\mp\pi^\pm$ can be defined in a similar way.
However, the penguin pollution $P$ is of $O(\lambda)$ relative to
$T$ in these decays, such that a larger deviation of $S_{\pi\pi}$
from $S_{c\bar cs}$ was found. The PQCD results of $S_{\pi\pi}$
are consistent with the data, but those of $S_{\pi^0K_S}$ are not.
Moreover, PQCD predicts $\Delta S_{\pi^0 K_S}> 0$, opposite to the
measured value. This result is in agreement with those derived in
the literature \cite{CGRS,CCS2,B05,WZ06}. Hence, it is not easy to
explain the data of $S_{\pi^0K_S}$ \cite{IDL}.

\begin{table}[ht]
\begin{tabular}{cccccccc}
\hline & Data & LO & LO$_{\rm NLOWC}$& +VC & +QL &  +MP  & +NLO
\\
\hline $S_{\pi^0 K_S}$ &
 $\phantom{-}0.31\pm0.26$&
 $\phantom{-}0.70$& $\phantom{-}0.73$& $\phantom{-}0.74$&
 $\phantom{-}0.73$& $\phantom{-}0.73$&
 $\phantom{-}0.74^{+0.02\,(+0.01)}_{-0.03\,(-0.01)}$
\\
$S_{\pi\pi}$ &$ - 0.50 \pm 0.12 $ &
 $-0.34$&$-0.49$&$-0.47$&$-0.51$&$-0.41$&
 $-0.42^{+1.00\,(+0.05)}_{-0.56\,(-0.05)}$
\\\hline
\end{tabular}
\caption{Mixing-induced CP asymmetries from PQCD in the NDR
scheme.
}\label{mixcp}
\end{table}

\section{Conclusion}

Many puzzles in exclusive $B$ meson decays have been observed
recently. The data $A_{CP}(B^\pm\to \pi^0 K^\pm)$ much different
from $A_{CP}(B^0\to \pi^\mp K^\pm)$ could be resolved in NLO PQCD
by taking into account the vertex corrections. We found that there
is no satisfactory resolution to the $B\to\pi\pi$ puzzle in the
literature: the available proposals are either data fitting, or
can not survive the constraints from the $B\to\rho\rho$ data under
the current theoretical development. The NLO effects push the
deviation $\Delta S_{\pi^0 K_S}$ toward the even larger positive
value. Therefore, the measurement of the mixing-induced CP
asymmetries in the penguin-dominated modes provides an opportunity
of discovering new physics.

\vskip 1.0cm

This work was supported by the National Science Council of R.O.C.
under Grant No. NSC-94-2112-M-001-001, and by the Taipei branch of
the National Center for Theoretical Sciences. I thank N. Sinha for
her hospitality during the Whepp9 workshop.

\end{document}